\documentclass[pre,twocolumn,aps,showpacs,floatfix]{revtex4}

\usepackage{graphicx}
\usepackage{amsmath}
\usepackage{amsfonts}
\usepackage{amssymb}
\usepackage{epsf}
\usepackage{bm}

\begin{document}

\title{Dispersionless transport in a washboard potential}

\author{Katja Lindenberg$^{1}$,
J. M. Sancho$^{2}$, A. M. Lacasta$^{3}$, and
I. M. Sokolov$^{4}$
}
\affiliation{
$^{(1)}$
Department of Chemistry and Biochemistry 0340, and Institute for
Nonlinear Science,
University of California, San Diego, La Jolla, California 92093-0340,
USA\\
$^{(2)}$
Departament d'Estructura i Constituents de la Mat\`eria,
Facultat de F\'{\i}sica, Universitat de Barcelona,
Diagonal 647, E-08028 Barcelona, Spain\\
$^{(3)}$
Departament de F\'{\i}sica Aplicada,
Universitat Polit\`{e}cnica de Catalunya,
Avinguda Doctor Mara\~{n}on 44, E-08028 Barcelona, Spain\\
$^{(4)}$
Institut f\"{u}r Physik, Humboldt Universit\"{a}t zu Berlin,
Newtonstr. 15, 12489 Berlin, Germany
}

\date{\today}

\begin{abstract}

We study and characterize a new dynamical regime of underdamped
particles in a tilted washboard potential. 
We find that for small friction in a finite  range of forces 
the particles move essentially nondispersively, that is, coherently,
over long intervals of time.  The associated distribution of the 
particle positions moves at an essentially constant velocity
and is far from Gaussian-like.  This new regime is complementary to,
and entirely different from, well-known nonlinear response and
large dispersion regimes observed for other values of the external
force. 
\end{abstract} 


\maketitle

Particle transport and diffusion in periodic potentials
at finite temperatures has been addressed in so many contexts and over so
many decades that one might think this to be a fully
solved problem~\cite{risken}.
However,
as modern experimental and numerical methods
continually evolve, ever broader parameter regimes and time regimes
become accessible to inquiry, and behaviors continue to be revealed
that have not previously been explored or even
noted~\cite{marchesoni,reimann,heinsalu,hayashi,tatarkova,coffey,machura,njp,we}. 
Intermediate time regimes are especially challenging.  On the
one hand, numerical methods need to be efficient to reach beyond 
relatively short time behavior.  On the other, analytic methods usually
deal with asymptotia. Yet, experiments often
involve intermediate time regimes.  In this Letter we
report an unexplored dynamical regime, namely, particle
transport that is essentially \emph{nondispersive} or \emph{coherent}
over long time intervals.  

Consider a particle moving in a
periodic potential $V(x)$ of amplitude $V_0$ and period $\lambda$,
with coefficient of friction $\eta$ at temperature $T$,
and subject to a constant external force $f$.
The variables $x$ and $t$ respectively denote the position of the
particle and the time. 
The equation of motion reads
\begin{equation}
\ddot{r}
= -{\cal {V}}^\prime (r)
-\gamma\dot{r}+F+\zeta(\tau),
\label{scaledequation}
\end{equation}
where $r=x/\lambda$, $\tau=(V_0/m)^{1/2}t/\lambda$, 
the dot and prime denote derivatives with respect to
$\tau$ and $r$ respectively, ${\cal{V}}(r)= V(x)/V_0$,
and the
noise obeys the fluctuation-dissipation relation
$\langle \zeta(\tau)\zeta(\tau')\rangle =2\gamma {\mathcal T}
\delta(\tau-\tau')$.
Equation~(\ref{scaledequation}) models the translational Brownian motion
of a particle in a tilted periodic potential, and also the rotational
Brownian motion of a damped pendulum driven by a constant torque. The
pendulum provides the mathematical background underlying a number of
applications, including mobility in superionic conductors, dynamics of
charge-density waves, ring laser gyroscopes, and phase-locking phenomena
in radio engineering.  Perhaps most directly relevant for experimental
testing, it also models a resistively and capacitively shunted single
Josephson junction.  This latter correspondence 
has been invoked in some of the most recent work on transport in
tilted periodic potentials~\cite{coffey,machura}.
An excellent table indicating the
precise translation between the parameters of
a number of physical systems including the quintessential Josephson
junction testbed and our model equation can be found in~\cite{coffey}. 

There are
three independent
parameters in the model, the scaled force $F = \lambda f/V_0$,
the scaled temperature ${\cal T} = k_BT/V_0$, and the scaled
dissipation $\gamma = \eta \lambda/(mV_0)^{1/2}$.
In our subsequent numerical simulations we choose 
${\cal V}(r) = -(1/2) \cos (2\pi r)$, $\gamma=0.04$, 
and ${\mathcal T}=0.2$.

\begin{figure}
\begin{center}
\includegraphics[height = 6cm]{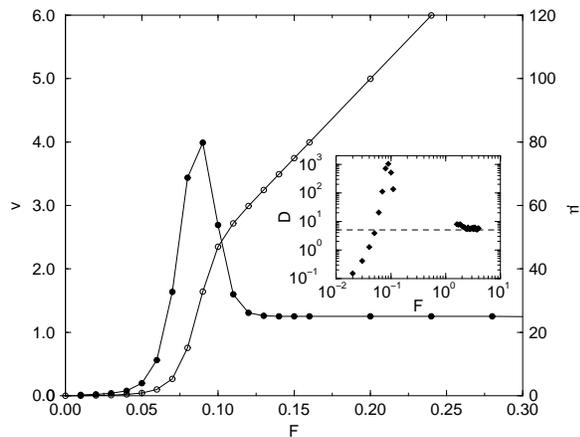}
\end{center}
\vspace*{-0.2in}
\caption{
Mean velocity $v$ (left axis, open circles) and mobility $\mu$
(right axis, black circles) vs the external force. 
Inset: diffusion coefficient $D$
vs external force. The ``empty" portion of the
diffusion coefficient curve indicates that we did not achieve a
stationary state up to the longest simulation times ($\tau\sim
O(10^7)$). This is the regime of persistent dispersionless transport
(see text).
}
\label{fig1}
\end{figure}

The external force $F$ (which we choose positive) tilts
the potential, ${\cal V}(r) -Fr$. 
For small forces, the tilted potential 
has wells and barriers (``washboard potential").  Beyond a critical
force $F_c = \pi$ the maxima and minima disappear, and the
potential decreases monotonically with increasing $r$. 
For small $F$ the motion of a particle is
essentially confined to the potential wells from which a thermal
fluctuation occasionally causes it to emerge. The velocity and dispersion
of the particles in this ``locked state" can be understood from
well-known Kramers arguments. 
On the other hand, when $F>F_c$,
the periodic portion of the potential becomes unimportant;
the particles
follow the force with an average velocity
that approaches $v=F/\gamma$ with increasing force,
with a dispersion given by the Einstein law of diffusion.  This is the
``running state." When the force is not too low but still
well below $F_c$, there is a regime where particles readily alternate
between the locked and running states.  As a result, the mean velocity of
the particles rises rapidly with increasing force,
the mobility exhibits a peak associated with this rapid rise, and the
dispersion also exhibits a peak that reflects the bimodal
distribution of velocities~\cite{risken,njp}.  These well-known behaviors
for underdamped particles are
illustrated in Fig.~\ref{fig1},
where we show the
asymptotic mean velocity
$v=\lim_{\tau\to\infty} \langle r(\tau)\rangle/\tau$
of the particles as a function of $F$,
and the associated mobility $\mu=dv/dF$. In the inset we display
the diffusion coefficient 
$D=\lim_{\tau\to\infty} \langle\Delta r^2(\tau)\rangle/2\tau$, where
$\langle\Delta r^2(\tau)\rangle
= \langle[r(\tau)-\langle r(\tau)\rangle]^2\rangle$
is the relative mean square displacement or dispersion.
%
The mobility and diffusion coefficient peak at a
$\gamma$-dependent force more than an order of magnitude below $F_c$ for
$\gamma=0.04$.  Beyond the peaking force, the mobility 
reaches a constant
close to $1/\gamma$.  The mean velocity
rises quickly as particles emerge from the well, and settles
to its steady state value in times of $O(10^2-10^3)$
(explicitly estimated later) for our parameters.

The as yet unexplored behavior occurs for forces
above those of the mobility and diffusion peaks but below the critical
force.  In this regime, which
extends from well below up to nearly $F_c$, we have
exhibited no values for $D$ in the inset of
Fig.~\ref{fig1}.  This is because for
times that are as long as we are able to reasonably run our simulations 
($\tau \sim 10^7$), $\langle \Delta r^2\rangle$
has not yet reached a behavior proportional to time.
For this wide range of intermediate forces $\langle \Delta r^2 \rangle$
remains essentially \emph{constant} over time intervals
that can extend over several decades, 
i.e., \emph{the transport is seemingly dispersionless}. This
coherent behavior occurs only in the underdamped system and requires 
there to be a washboard potential. 
The nondispersive behavior is remarkable because it occurs in the
presence of thermal fluctuations.
At sufficiently long times $\langle \Delta r^2 \rangle$
begins to grow linearly in time, presumably also for those cases in
which we were not able to reach this regime.

\begin{figure}
\begin{center}
\includegraphics[height = 6cm]{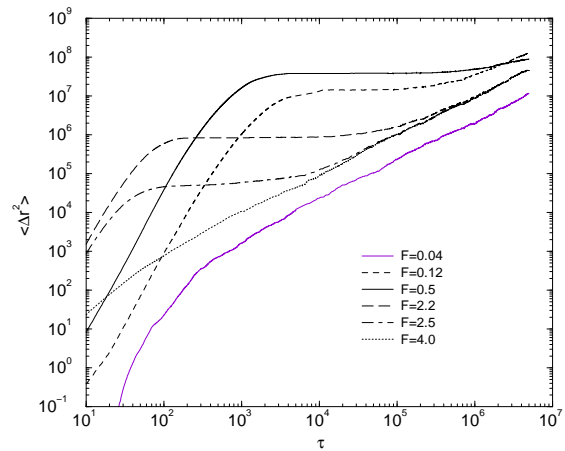}
\end{center}
\vspace*{-0.2in}
\caption{Dispersion vs time for different forces. 
Note that the curves for small ($F=0.04$) and 
large ($F=4.0$) forces quickly cross over from the short-time
superdiffusive regime typical of low friction to the asymptotic
diffusive regime. Intermediate forces exhibit a long dispersionless
regime, eventually followed by normal dispersion.}
\label{fig2}
\end{figure}

Figure~\ref{fig2} shows the dispersion
for different values of the force. 
For very small forces (including zero force), and also for very
large forces, $\langle \Delta r^2 \rangle$
settles into the expected linear growth with time on the same time
scale as the mean velocity. The corresponding diffusion coefficients
as reflected in the linear growth slopes are those shown in the inset
of Fig.~\ref{fig1}.  In Fig.~\ref{fig2} we show two
forces ($F=2.2$ and $2.5$) which exhibit the
full range of behaviors of the dispersion from early rise to
dispersionless regime followed by normal Einstein behavior. 
We also show an example of a
force, $F=0.5$, for which the dispersionless regime extends over
almost the entire time range of the figure after the initial rise. 

\begin{figure}
\begin{center}
\includegraphics[width = 8cm]{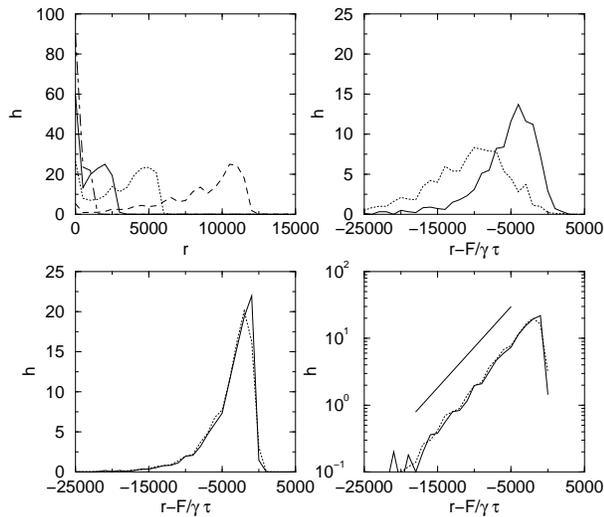}
\end{center}
\vspace*{-0.2in}
\caption{Unnormalized distribution of particle positions at different
times for $F=0.12$.  Top left: early times regime. Values of $\tau$:
$500$ (dot-dashed); $1000$ (solid); $2000$ (dotted); $4000$ (dashed).
Bottom left: dispersionless time regime for 
$\tau=7000$ (solid), $30,000$ (dotted). Top right: long-time dispersive
regime for $\tau=2\times 10^5$ (solid), $10^6$ (dotted). 
Bottom right: Log plot of bottom left distribution to display the
exponential tail. The straight line is the exponential form~(\ref{distrib})
with $l_0=(F/\gamma)\tau_0=3600$. 
}
\label{fig3}
\end{figure}

To explore the nondispersive behavior in more detail, 
in Fig~\ref{fig3} we display simulation results
for the
distribution of the moving particles.
Initially all the particles are placed
at the minimum of one potential well, with a
Maxwell-Boltzmann distribution of velocities
at temperature ${\mathcal T}$.  At early times (upper left panel)
the distribution has a large peak around the origin because most
of the particles are still in the initial well.  As time proceeds, 
particles
leave the well, giving rise to the second peak.
The growing dispersion in this regime is due to the bimodal 
distribution. 
The maximum of the second peak
moves practically from the very beginning with velocity
$v\approx F/\gamma$.  In time, the first peak
disappears and the second peak settles into the behavior shown in
the lower panels, which display the
remarkable intermediate-time nondispersive behavior.
The distribution of positions is a one-sided exponential
that moves coherently without distortion or dispersion over a long
time interval.  This is particularly remarkable since
during all this time (as well as before and after)
the particles are persistently subject to
thermal fluctuations.  At sufficiently long times, dispersion resumes and
the distribution not only broadens but approaches the
characteristic equilibrium Gaussian shape (upper right panel).
We have ascertained this same behavior for other forces.
We proceed to identify the physical mechanism
and to estimate the forces and
crossover times for the occurrence of the dispersionless behavior.

Initially all the particles are placed
in one potential well, from
which they emerge according to the well-known exponential distribution
of first exit times $\tau_e$,
$P(\tau_e)= \tau_0^{-1} \exp(-\tau_e/\tau_0)$~\cite{redner}.
The mean exit time $\tau_0$ depends on the system
parameters, the temperature, and the external force.
Since the particles emerge with velocities very narrowly distributed
about $v\approx F/\gamma$,
this temporal distribution
can be transformed to a spatial distribution
of particles at $\tau=0$, $p(l_e,0)= l_0^{-1}\exp(l_e/l_0),~{-\infty}
<l_e\leq 0$, via
the change of variables $l_e=(F/\gamma)\tau_e$. 
Here $l_0\equiv l_{\tau_0}\equiv (F/\gamma)\tau_0$.
This distribution looks precisely like the ones in
the lower panels of Fig.~\ref{fig3}. 
At time $\tau$ the distribution will have moved to the right 
pretty much undistorted, 
\begin{equation}
p(l_e,\tau) = \frac{1}{l_0} \exp\left(\frac{l_e-l_\tau}{l_0}\right),
\qquad l_\tau\equiv (F/\gamma)\tau.
\label{distrib}
\end{equation}
The moments of this distribution are consistent with the
numerical results. In particular, when $l_\tau\gg l_0$ the first moment
$\to l_\tau$.  The dispersion settles to the \emph{constant} value
$l_0^2$.  These results reproduce the intermediate time behaviors
highlighted in the simulation results. 

The parameter $\tau_0$ defines the crossover time
from short to intermediate time behavior.  Its value 
can be extracted from the simulation results.  We find
$\tau_0=1200~(F=0.12)$, $490~(F=0.5)$, $17~(F=2.2)$, $3.8~(F=2.5)$.
A theoretical estimate
follows
from the
well-known result for the escape time of an underdamped
particle from a potential well~\cite{melnikov,hanggi,nosotros},
$\tau_0 \propto ({\mathcal T}/\gamma U)\exp{U/{\mathcal T}}$.
The proportionality factor varies
depending on
the precise definition of $\tau_0$, but is generally of $O(1)$. 
For small $F$, $U=1-F/2 +O(F^2)$; for larger $F$, $U$ is determined from a
transcendental equation.  This estimate leads to results consistent
with the simulation values.

What marks the existence and termination of the dispersionless transport
regime? There are two contributions to the width of 
$p(l_e,\tau)$.  One is the width $l_0$ due to
dispersion of emergence times from the
potential well, and the other comes from the
thermal fluctuations.  The total square width of the distribution is obtained
from a convolution of these two contributions and is thus simply their
sum, $\langle \Delta r^2(\tau)\rangle = l_0^2 +2D\tau$. We stress that
$D$ is the actual diffusion coefficient, which is in general different from
${\mathcal T/\gamma}$ (see inset in Fig.~\ref{fig1}).  The effects of
the thermal fluctuations are therefore submerged until times $\tau\sim
l_0^2/2D$. This estimate leads to the dispersionless ending times
$\tau=5.4\times 10^6$ ($F=0.12$),
$8.9\times 10^4$ ($F=2.2$),
and $5.6\times 10^3$ ($F=2.5$), consistent with the results in
Fig.~\ref{fig2}.  The seemingly anomalous dispersionless behavior is
thus a \emph{natural consequence} of the motion of an underdamped
particle in a washboard potential. 

A necessary condition  for nondispersive motion
is that the motion of a single particle be nonchaotic in the absence of
noise. 
An upper bound on the forces that lead to
nondispersive motion is $F=F_c$.  A lower bound can be
estimated from Fig.~\ref{fig1} as the force above which $v\approx
F/\gamma$, which for $\gamma=0.04$ is around $F\sim 0.1$. To estimate
this bound analytically and, more generally, to
analyze the behavior of single particle
trajectories through the dispersionless and into the dispersive regimes,
we define the auxiliary variable $z(\tau)$ that
describes the deviations of the particle trajectory from the approximate
mean trajectory, $z = r -[l_\tau-l_e]$.  The distance $l_e$
quickly becomes negligible.  The equation of evolution for $z(\tau)$ is
$\ddot{z} = -\pi \sin 2\pi(z+l_\tau) -\gamma\dot{z}+ \zeta(\tau)$,
which describes a particle moving in a cosinusoidal
potential that is itself rapidly and periodically
oscillating in time.  In the absence of noise,
suppose that this particle is near the bottom of a well, 
slowly relaxing toward the minimum.
At a time $\gamma/2F$ later it finds itself near
the top of the potential, from which the small damping slowly pulls it toward
the nearest minimum (either the same it was in or an adjacent one,
depending on its precise location).  
Before it has gone far the particle again finds
itself near the minimum of the well as the potential flips,
the very same minimum in which it started.  Again the damping pulls it
toward the bottom, and the cycle restarts.  The
resulting $z(t)$ is thus expected to remain very small.
An approximate solution for small displacements
in the absence of the noise is easily obtained 
by approximating the nonlinear problem as a dichotomous succession of
linearized problems,
$\ddot{z} = -(+)2\pi^2 z -\gamma\dot{z}$ for $\tau_{2n} < \tau < \tau_{2n+1}$
($\tau_{2n+1}< \tau <\tau_{2n+2}$),
where $\tau_n\equiv nF/2\gamma$ and $n=0,1,2,\cdots$.
The solution of the two equations is straightforward, particularly at
times $t_{2n}$, 
\begin{equation}
z(t_{2n}) = a e^{(-n\gamma^2/2F)}
\cos \left (\theta + n \arctan
\frac{\sqrt{1-A^2}}{A}\right),
\label{zagain}
\end{equation}
where $A\equiv \cosh u \cos u$,
$u=\pi\gamma/\sqrt{2}F$, $z(0) = a\cos\theta$, and we have retained only
leading contributions in $\gamma/F$. 
The switching thus leads
to a decay of solutions down to zero, which is the stabilizing behavior
that leads to dispersionless motion.  
The solution~(\ref{zagain}) holds when
$A^2<1$,
which places the constraint $F> 1.2\gamma$
on the parameters.  For the value $\gamma=0.04$ used in our
simulations, $F>0.05$, in
gratifying agreement with the lowest forces found to
lead to dispersionless transport. 

\begin{figure}
\begin{center}
\includegraphics[height = 5cm,width=8cm]{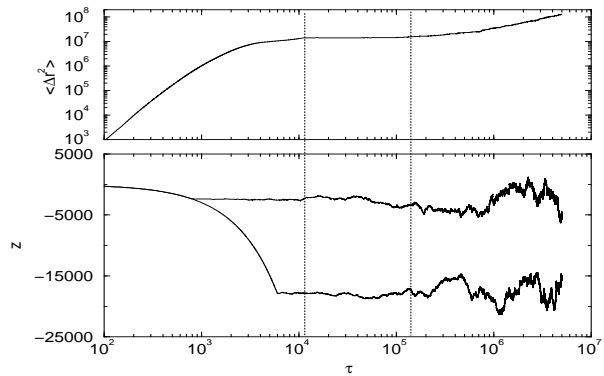}
\end{center}
\vspace*{-0.2in}
\caption{
Bottom panel: Two individual $z$-trajectories for $F=0.12$. 
The two trajectories
emerge from the initial potential well at slightly different times of
order $10^3$ and move essentially nondispersively for a long time, until
dispersive motion eventually sets in.  Upper panel: associated
dispersion curve.  
}
\label{fig4}
\end{figure}

A clear manifestation of this description can be seen in
Fig.~\ref{fig4}. 
We show two individual $z$-trajectories
that emerge from the potential well at different times.
The associated particle trajectories are therefore in different
portions of the exponential distribution of Fig.~\ref{fig3}.
The remarkable point is the flat $z$-trajectory seen in each case in
the time interval between its emergence from the well to around
$\tau\gtrsim 10^5$.
This is the regime of dispersionless motion, 
as seen in the associated dispersion curve in the upper panel.

These arguments do not include the effects of noise, which buffets $z$
and eventually causes it to increase beyond
the validity of our linearized theory. The system
then no longer finds itself near the top and bottom of the flip-flopping
potential but rather is sufficiently out of phase with the extrema 
so that the alternation does not
repeatedly drive it back to the same minimum.  Dispersion then sets in.
Figure~\ref{fig4} confirms that
the dispersionless motion
comes to an end, and noisy motion associated with linear
growth of $\langle \Delta r^2\rangle$ with time resumes.

In conclusion, we have identified and characterized a transport regime in
the motion of thermally agitated classical particles 
that involves essentially dispersionless motion over several decades
of time in appropriate parameter regimes.  This remarkably coherent
behavior requires the presence of thermal fluctuations, and is
restricted to underdamped systems in
a washboard potential.
We have provided a theoretical framework to
estimate the times of onset and duration of dispersionless transport
and the parameter regimes in which it occurs.  Eventually,
dispersive motion sets in again and the system resumes the
familiar behavior, but it may take an inordinately long time to do so,
so long that it lies beyond our computational means. 
We are confident that the results obtained here are amenable to
experimental verification.  Apart from the possibility of a mechanical
or electronic realization of a tilted washboard system, the regime might
be observed in Josephson junctions, where the voltage (the observable) is
proportional to the time derivative of the phase (corresponding to the
particle coordinate in our model).
It is important to note that we have also observed nondispersive behavior
in numerical simulations in two dimensions with a separable potential of
the form ${\cal V}(r_x,r_y) = -(1/2) [\cos (2\pi r_x) + \cos (2\pi
r_y)]$.  This suggests that our results may be testable on the motion of
molecules and clusters on surfaces [see references in~\cite{we}].
These observations
give us reason to expect that this coherent dispersionless behavior
should be experimentally accessible. 

This work was partially supported by the National Science Foundation
under Grant No. PHY-0354937 (KL), by the MCyT (Spain) under project
FIS2006-11452 (JMS and AML), and by the HPC-Europa Transnational Access
Program (IMS).


\begin{thebibliography}{99}
\bibitem{risken}  
H. Risken, \emph{The Fokker-Planck Equation} (Springer Verlag,New York,
1989), Ch. 11.

\bibitem{marchesoni} F. Marchesoni, Phys. Lett. A {\bf 231}, 61
(1997); G. Costantini and F. Marchesoni, Europhys. Lett. 
{\bf 48}, 491 (1999); M. Borromeo and F. Marchesoni, Surf. Sci. {\bf 
465}, L771 (2000).


\bibitem{reimann}
P. Reimann  {\em et al.}, Phys. Rev. Lett. {\bf 87}, 
010602 (2001); P. Reimann  {\em et al.}, Phys. Rev. E {\bf 65},
031104 (2002).

\bibitem{heinsalu} E. Heinsalu, R. Tammelo, and T. \"{O}rg, Phys.
Rev E {\bf 69}, 021111 (2004).

\bibitem{hayashi} K. Hayashi and S. Sas, Phys. Rev. E {\bf 69},
066119 (2004).

\bibitem{tatarkova} S. A. Tatarkova, W. Sibbett, and K. Dholakia,
Phys. Rev. Lett. {\bf 91}, 038101 (2003).

\bibitem{coffey} W. T. Coffey, Yu. P. Kalmykov, S. V. Titov, and B. P.
Mulligan, Phys. Rev. e {\bf 73}, 061101 (2006).

\bibitem{machura} L. Machura, M. Kostur, P. Talkner, J. Luczka, and P.
H\"anggi, cond-mat/0609452.

%
%
%
%

\bibitem{njp} K. Lindenberg, A. M. Lacasta, J. M. Sancho, and A. H.
Romero, New J. Phys. {\bf 7}, 29 (2005).

\bibitem{we} J. M. Sancho, A. M. Lacasta, K. Lindenberg, I. M. Sokolov,
and A. H. Romero, Phys. Rev. Lett. {\bf 92}, 250601 (2004); 
A. M. Lacasta, J. M. Sancho, A. H. Romero, I. M. Sokolov,
and K. Lindenberg, Phys. Rev. E {\bf 70}, 051104 (2004).

\bibitem{redner}
S. Redner, {\em A Guide to First-Passage Processes} (Cambridge
University Press, 2001).

\bibitem{melnikov}
V. I. Melnikov and S. V. Meshkov, J. Chem. Phys. {\bf 91}, 4073 (1986);
V. I. Melnikov, Phys. Rep. {\bf 209}, 1 (1991).

\bibitem{hanggi}
P. H\"anggi, P. Talkner, and M. Borkovec, Rev. Mod. Phys. {\bf 62}, 251
(1990).


\bibitem{nosotros}
J. M. Sancho, A. H. Romero, and K. Lindenberg, J. Chem. Phys. {\bf 109},
9888 (1998).

\end{thebibliography}
\end{document}